\documentclass{ws-ijmpc}

\begin{document}

\markboth{M. Gmitra, D. Horv\'ath}
{Properties of iterative Monte Carlo single histogram reweighting}

\catchline{}{}{}{}{}

\title{PROPERTIES OF ITERATIVE MONTE CARLO SINGLE HISTOGRAM REWEIGHTING}

\author{MARTIN GMITRA}

\address{ Department of Theoretical Physics and Astrophysics,
          P.J.\v{S}af\'arik University,\\
          Park Angelinum 9, 040 01 Ko\v{s}ice, Slovak Republic \\
          gmitra@kosice.upjs.sk}

\author{DENIS HORV\'ATH}

\address{ Department of Theoretical Physics and Astrophysics,
          P.J.\v{S}af\'arik University,\\
          Park Angelinum 9, 040 01 Ko\v{s}ice, Slovak Republic \\
          horvath.denis@gmail.com}

\maketitle

\begin{history}
\received{Day Month Year}
\revised{Day Month Year}
\end{history}

\begin{abstract}
We present iterative Monte Carlo algorithm
for which the temperature variable
is attracted by a critical point.
The algorithm combines techniques
of single histogram reweighting and
linear filtering. The 2d~Ising model of ferromagnet
is studied numerically as an illustration.
In that case, the iterations uncovered
stationary regime with invariant probability
distribution function of temperature
which is peaked nearly the pseudocritical
temperature of specific heat.
The sequence of generated temperatures
is analyzed in terms of stochastic
autoregressive model. The error of
histogram reweighting can be better understood
within the suggested model. The presented model
yields a simple relation, connecting variance 
of pseudocritical temperature
and parameter of linear filtering.

\end{abstract}

\ccode{PACS Nos.: 02.70.Tt, 05.50.+q, 02.50.Ey, 05.65.+b}

\section{Introduction}
The critical phenomena, typical of many body systems,
have attracted remarkable scientific interest since a long time.
The Monte Carlo~(MC) stochastic techniques~\cite{Binder1988}
serve to estimate thermodynamic averages from samples of the 
configuration space. Many advanced simulation approaches
are rooted in elementary updates of Metropolis algorithm~\cite{Metropolis},
which generates a Markovian chain of spin configurations.

Most of the efforts try to improve MC method
within the critical region. The efficiency of MC method have been 
increased by the development of sampling and data processing techniques.
The reweighting technique applied anew to classical statistical
systems~\cite{his_reweight,Ferrenberg63} enables to perform a random 
sampling for one distribution function, and to calculate
quantities of interest for similar distributions.
To be more concrete, let us assume that a sample of $N$ configurations
($q=1,2,\ldots, N$) of quantity $Q$ have been accumulated
for a fixed temperature $T_t$.
The samples $Q_q$ are used to construct a continuous temperature
$T$ dependence of a single histogram approximation 
$\langle Q \rangle_{T,T_t}$ of the canonical
thermodynamic average $\langle Q \rangle_T$.
The interpolation called "{\em reweighting
on the fly}"~\cite{Ferrenberg95} is based on the formula
\begin{equation}
\langle Q \rangle_{T,T_t} =
\frac
{\displaystyle \sum_{q=1}^{N} Q_q \exp \left[ \frac{E_q}{T_t}
  \left( 1 - \frac{T_t}{T}\right) \right]}
{\displaystyle \sum_{q=1}^{N}
\exp \left[ \frac{E_q}{T_t}\left(1 - \frac{T_t}{T} \right) \right]}\,,
\label{Reweight1}
\end{equation}
where $E_q$ is the energy~(in Boltzmann constant units)
of configuration $q$. Because of the finite length
of the MC run, Eq.~(\ref{Reweight1}) provides reliable results
only for a relatively narrow range of $T$ values around $T_t$.
When $T$ differs too much from $T_t$, it causes an increase
of the statistical errors and may lead to unreliable results.
The reliable range of $T$ values also decreases as the system size
increases~\cite{Ferrenberg91,Landau99}.
A qualitative study in Ref.~\refcite{Munger91} has shown that
reweighting causes a systematic shift in the height and the location 
of the specific heat peak, which depends on the relative position of $T_t$ and $T$.
It turns out that the shift decreases as~$\sim N^{-1/2}$
only for $N$ much larger than the number of degrees of freedom.
Systematic errors, eventually affected by sampling
autocorrelations, may also take place at finite $N$.
The first quantitative study of the statistical errors in
reweighted data~\cite{Ferrenberg95} has demonstrated that
theoretical calculation of the error is both difficult and
time consuming so it may not be justified for all studies.
In this paper we present rather general iterative treatment based on the
single histogram reweighting which implicitly takes into
account both the systematic errors of reweighting and the statistical
errors to generate stochastic fluctuations near a critical point.

A well known systematic error, present in the histogram method,
is an overestimation (underestimation) of $\langle E \rangle_{T,T_t}$
when $T \ll T_t$ ($T \gg T_t$), respectively~\cite{Ferrenberg91}.
This fact and the general idea of feedback regulation~\cite{Kadanoff}
are used to construct an iterative algorithm.
The aim of the suggested method is to estimate the critical temperature 
or other parameters of interest, as well as to perform an analysis 
and partial elimination of their errors. The method combines single 
reweighting, linear filtering and an iterative treatment.
The incorporation of iterations is inspired by a general
class of probabilistic models~\cite{Robbins51}.
The iterations generate data that are further analyzed
in terms of a suggested autoregressive moving average~(ARMA) model.

The single-spin flip Metropolis algorithm is applied here
in order to obtain averages for moderate lattice sizes.
The cluster methods are not used due to the efficiency
reasons discussed in~Ref.~\refcite{Ito94}. The efficiency of the algorithm
should also be increased by hybrid schemes~\cite{Plascak02}.
However, because of many side effects that can show up, a separate study
of hybrid schemes is needed.

The paper is organized as follows. In Sec.~\ref{general_scheme}
the iterative self-organized algorithm is suggested.
In Sec.~\ref{sec:model} the autoregressive phenomenological model
of the stochastic iterative search process is proposed 
and the corresponding relations for second-order statistics
are derived. The model is used to parametrize
MC data obtained for ferromagnetic 2d~Ising spin models on $L\times L$
square lattices~(see Sec.~\ref{sec:details}).

\section{The iterations near the criticality}
\label{general_scheme}
We start by defining a process where temperature is driven 
by some response function. The stochastic process can
be formally described by the second-order recursive formula
\begin{eqnarray}
T^{\rm his}_{t+1}  &=& F_{T_t,N}\,,
\label{EqRecur1}
\\
T_{t+1} &=&  \eta
T^{\rm his}_{t+1} + (1-\eta) T_t\,.
\label{Adap1}
\end{eqnarray}
It generates sequence
$\{\, T_t, T_{t=0}=T_0,t=1, 2, \ldots, N_{\rm event}\}$.
The stochastic function $F:\, T_t \rightarrow T^{\rm his}_{t+1}$,
which provides a {\em preliminary histogrammatic}
estimate $T^{\rm his}_{t+1}$ of unknown critical temperature $T^{\ast}$,
is defined implicitly by
\begin{equation}
F:\,\,\,\, C_{T^{\rm his}_{t+1},T_t,N}
=  \max_{T \in \langle 0,\infty \rangle}
C_{T,T_t,N}\,.
\label{Objec1}
\end{equation}
The key point of the reweighting is to localize the extremal value
of the response function $C$ evaluated for $N$ MC samples.
Commonly, the response function can be defined by the first and second
moments of energy or magnetization, respectively.
The role of filtering, defined by the plasticity parameter 
$\eta$ in Eq.~(\ref{Adap1}), is to weaken fluctuations~\cite{Principe2000} 
of the $T^{\rm his}_{t+1}$ sequence.

Let us assume the invariant probability distribution function~(pdf)
of the form
\begin{equation}
p(T)=\overline{\delta_{T,T_t}} =
\lim_{t_{\rm max}\rightarrow \infty}
\frac{1}{t_{\rm max}}\,
\sum_{t=1}^{t_{\rm max}} \delta_{T,T_{t}}\,\,\,,
\label{Ttx12}
\end{equation}
where $\delta$ is the Kronecker's symbol and
$\overline{\cdots}$~is the arithmetic average~(which differs
from standard thermodynamic average $\langle\ldots\rangle$).
From now on $T^{\ast}$ is associated with pseudocritical temperature
$T_{\rm c}(L)$ of the specific heat. For Gaussian-like $p(T)$ we suppose
\begin{eqnarray}
T^{\ast} &\simeq &
\lim_{t_{\rm max}\rightarrow\infty}
\frac{1}{t_{\rm max}}
\sum_{t=0}^{t_{\rm max}} T_t =
\overline{T_t}
\label{Tavel1}
\\
&=&
\int_0^{\infty} \,
{\rm d}T \, p(T)\,T\, .
\nonumber
\end{eqnarray}
As mentioned above, the specific heat can be estimated within 
a histogram reweighting approach using the fluctuation-dissipation relation
\begin{equation}
C_{T,T_t} =
\frac{\langle E^2 \rangle_{T,T_t} -
\langle E \rangle^2_{T,T_t}}{T^2 L^2}\,.
\label{Objec2}
\end{equation}
At the thermodynamic limit $N\rightarrow \infty$, the iterative process
becomes deterministic and tends to an unique fixed point
$T^{\ast}= F_{T^{\ast},N\rightarrow \infty}$.
For finite $N$, a convergence criterion have
to be formulated in a pure statistical sense.
The variance
\begin{equation}
{\cal V}_t=
\overline{  \left(\,T_t-T^{\ast}\, \right)^2 }
\label{Disp1}
\end{equation}
has been chosen as a suitable measure that reflects the convergence.

\section{The model of the iteration process}
\label{sec:model}

The iterative process, described by 
Eqs.~(\ref{EqRecur1}), (\ref{Adap1}), (\ref{Objec1})
supplemented by Eq.~(\ref{Objec2}), consists of nonlinear
and stochastic rules. As an auxiliary tool of analysis,
let us introduce a phenomenological model to parameterize the 
simulated statistics. Within the model, the explicit form 
of the stochastic function~[see Eq.~(\ref{EqRecur1})] is
\begin{equation}
F_{T_t,N}
=
\alpha
T^{\ast}
+
(1-\alpha) T_t +
\xi_t\,\,,
\label{model_alg}
\end{equation}
where an additive noise $\xi_t$ describes uncertainty.
The convergence rate towards $T^{\ast}$
is determined by the parameter $\alpha$,
which contains information of the systematic reweighting error.
From Eqs.~(\ref{EqRecur1}), (\ref{Adap1}) and Eq.~(\ref{model_alg}) 
follows that
\begin{equation}
T_{t+1} = \alpha\eta T^{\ast} + (1-\alpha \eta) T_t +  \eta \xi_t\,,
\label{model_2}
\end{equation}
where the noise term is controlled by the coefficient $\eta$.
The model discussed here can be considered
as an application of ARMA(1,1) model~\cite{IntroductionBrockwell1996}.
The solution of~Eq.~(\ref{model_2}) can be written simply as
\begin{equation}
T_t = T^{\ast} + (T_0-T^{\ast}) B^{-t} +
\eta B \sum_{q=0}^{t-1}\,
\xi_q B^{q-t} \,,
\label{soluTt1}
\end{equation}
where $B=1/(1-\alpha \eta)$.
The above formula, written for any $\xi_t$ dependence, is general.
At this point, let us consider the statistical properties
of Gaussian noise for $\xi_t$ specified by
\begin{eqnarray}
\overline{\,\xi_t\,} &=& 0\,\,,
\label{firstxi1}
\\
\overline{ \, \xi_t \, \xi_{t'}\,}\,&=& A\,
\delta_{t,t'}\,\,,
\label{Eq:whitenoise}
\end{eqnarray}
where $A>0$. Within the limit of a second order statistics in $T_t$,
from Eq.~(\ref{soluTt1}) and Eq.~(\ref{firstxi1}), follows
\begin{equation}
\overline{T_t} =
T^{\ast} + (T_0-T^{\ast}) B^{-t}\,.
\label{soluTt2}
\end{equation}
Therefore, the stationary regime is defined by the invariance
$\overline{T_t}=T^{\ast}$, which stays in agreement with Eq.~(\ref{Tavel1}).
If the solution given by Eq.~(\ref{soluTt1}) is substituted
into Eq.~(\ref{Disp1}), one obtains
\begin{equation}
{\cal V}_t =
B^{-2t} \left( T_0 - T^{\ast} \right)^2
+\frac{A \eta^2 B^2}{B^2-1}
\left(\,1 - B^{-2 t} \,\right)\, ,
\label{Disp2}
\end{equation}
which is equivalent to the recursion
${\cal V}_{t+1}=A\eta^2+{\cal V}_t\,B^{-2}$.
Terms proportional to $B^{-2t}$ describe a {\em transient regime} 
of the initial temperature $T_0$, which the system forgets
exponentially fast. Terms $(T_0 - T^{\ast})^2 B^{-2t}$ and
$A \eta^2 B^{2(1-t)}/(1-B^2)$ are deterministic and stochastic 
contribution to ${\cal V}_t$, respectively.
The variance~${\cal V}_t$ remains finite for
$t\rightarrow \infty$ if $B^2>1$, that is if
$0<\alpha \eta <2$. The transient regime can be characterized by
\begin{equation}
{\cal V}_1 =
A\eta^2 + ( 1-\alpha\eta )^2
( T_0 -T^{\ast} )^2\,\,,
\label{Eq:var_1}
\end{equation}
whereas, to obtain the stationary regime, one needs to calculate
\begin{equation}
{\cal V}_{t\rightarrow\infty} \equiv {\cal V}_{\infty}=
\frac{A  \eta^2 B^2}{B^2-1} = \frac{A\eta}{\alpha(2-\alpha\eta)}\,.
\label{Eq:var_infty}
\end{equation}
It results that ${\cal V}_{\infty}$ diverges at
$\alpha \rightarrow 2/\eta$ or equivalently $B\rightarrow -1$.
The characteristic transient time,
\begin{equation}
\tau_{\rm tr}=-\frac{1}{\ln|1-\alpha\eta|}\, ,
\label{time_ch}
\end{equation}
follows from its definition
$(B^2)^{-t} =\exp(-2t/\tau_{\rm tr})$.
The transient time has a singularity for $\eta\to (1/\alpha)^{\pm}$ and
$B\rightarrow \pm \infty$. In this limit the model describes immediate
convergence~($\tau_{\rm tr}\rightarrow 0$) and variance 
${\cal V}_{\infty} = A/\alpha^2$. A singular parametric
line, $1-\alpha \eta=0$, splits region $B^2>1$ into alternating~($B<-1$)
and unidirectional convergence~($B>1$) of $T_t$ near $T^{\ast}$.

Regarding the convergence properties, it is useful
to introduce $\eta=\eta_{\rm m}$ as the value of the
{\em highest initial convergence rate},
defined by the condition
${\rm d}{\cal V}_1/{\rm d}\eta|_{\eta=\eta_{\rm m}}=0$.
This yields
\begin{equation}
\label{Eq:eta_m}
\eta_{\rm m}
=\frac
{\displaystyle
\alpha\, \left(1-\frac{T_0}{T^{\ast}}\right)^2}
{\displaystyle
\frac{A}{(T^{\ast})^2}
+ \alpha^2
\left(\,1-
\frac{T_0}{T^{\ast}}\right)^2}\,.
\end{equation}
For this extremal value
is ${\cal V}_1 (\eta_{\rm m})= (A/{\alpha}) \eta_{\rm m}$.

Finally, we introduce some supplementary remarks about the
limit of the vanishing error. Clearly, for $N\rightarrow\infty$
we assume $\alpha\rightarrow 1^{-}$.
The consequence of Eqs.~(\ref{Eq:var_infty}) and (\ref{time_ch})
is $\eta \rightarrow 0^{+}$, that implies
${\cal V}_{t\rightarrow \infty}\rightarrow 0$.
Unfortunately, because of~$\tau_{\rm tr}\rightarrow\infty$
this cannot be attained in practice.

\section{Numerical results}
\label{sec:details}
In order to explore the proposed model by numerical simulation,
we chose the 2d~Ising model with the Metropolis algorithm as testing ground.
Most of the data are accumulated on square lattices of linear
dimension $L=10$ and periodic boundary conditions.
The methodology is expected to be quite general and of straightforward
application for both discrete and continuous systems.

Every iteration step $t=1,2,\ldots, N_{\rm event}$
of algorithm, based on Eq.~(\ref{EqRecur1}) and Eq.~(\ref{Adap1}),
consists of the following main blocks~(i)-(iv):
\begin{romanlist}[(ii)]
\item {\em equilibration}
that includes $10^4$ MCSS (Monte Carlo steps per spin) for constant $T_t$.
\item {\em accumulation}
of $N$ pairs $(Q_q, E_q)$ into histograms.
In order to reduce autocorrelations, successive samples are 
separated by one MCSS.
\item {\em localization}
of the maximum $C_{T^{\rm his}_{t+1},T_t,N}$ by a gradient algorithm.
Recall, that $C$ is computed from
Eq.~(\ref{Reweight1}) and Eq.~(\ref{Objec2}) for
$\langle E \rangle_{T,T_t}$ and $\langle E^2 \rangle_{T,T_t}$.
\item {\em prediction} of next $T_{t+1}$ using
the filtering given by Eq.~(\ref{Adap1}).
\end{romanlist}

\begin{figure}[ph]
\centerline{\epsfig{file=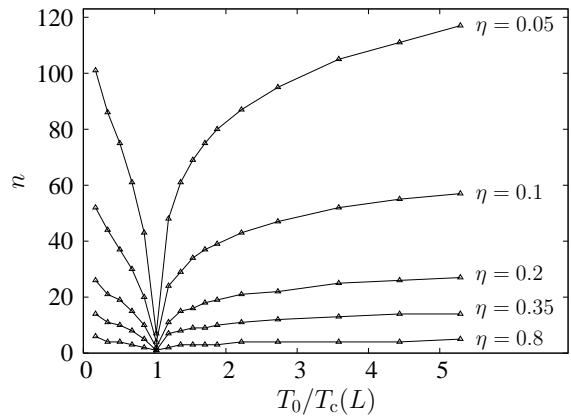,height=8cm,angle=-90.0}}
\caption{ The transient
          stages of the iterative regime
          characterized by the minimum number
          of iteration steps  $n$
          needed to satisfy
          the stop criterion
          $|\,T_n-T_{\rm c}(L)\,| \leq 0.01$
          for resulting
          $T^{\ast}=
          T_{\rm c}(L)=0.586141$.
          The number
          $n$ is plotted as a
          function of initial
          temperature
          $T_0$ [distinct
          from $T_{\rm c}(L)$]
          for several $\eta$.
          Calculated
          for fixed
          $N=10^5$.
\label{initconv1}}
\end{figure}

\begin{figure}[pt]
\centerline{\epsfig{file=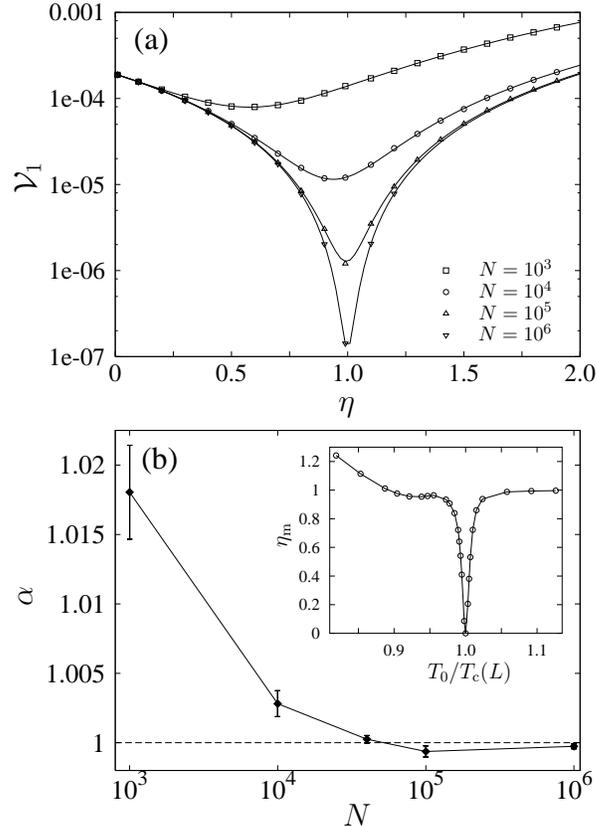,height=8cm,angle=-90.0}}
\caption{The analysis of transient regime:
   (a)~The $\eta$-dependence of variance
   ${\cal V}_1$ obtained for initial $T_0=0.6$.
   The lines correspond to the fits of $\eta$-dependences
   via Eq.~(\ref{Eq:var_1}). 
   The parameters are collected in Table~\ref{tab:par}.
   (b)~The $\alpha(N)$ dependence for $T_0=0.6$.
   The inset depicts $\eta_{\rm m}(T_0)$ for $N=10^4$.
\label{fig:var_1}}
\end{figure}

Fig.~\ref{initconv1}~illustrates the properties of
{\em transient regimes}, which are affected by the choice of
$T_0$ and $\eta$. The $\eta$-dependences of ${\cal V}_1$ 
are presented in~Fig.~\ref{fig:var_1}(a).
One may see that the agreement of the simulated data
with Eq.~(\ref{Eq:var_1}) is quite satisfactory.
Typical values of the parameters determined by
the fit are gathered in Table~\ref{tab:par}.

\begin{table}[ht]
\tbl{The transient regime parameters for $T_0=0.6$ 
obtained by fitting of Eq.~(\ref{Eq:var_1}).}
{\begin{tabular}{@{}cccc@{}}
$N$ & $\alpha$ & $A$ & $\eta_{\rm m}$\\
\hline \hline
$10^3$ & $1.01805$ & $1.398\cdot 10^{-4}$ & $0.5770$ \\
$10^4$ & $1.00281$ & $1.227\cdot 10^{-5}$ & $0.9376$ \\
$10^5$ & $0.99937$ & $1.282\cdot 10^{-6}$ & $0.9939$ \\
$10^6$ & $0.99974$ & $1.216\cdot 10^{-7}$ & $0.9996$ \\
\end{tabular} \label{tab:par}}
\end{table}

For~$N\to\infty$, we observe that $\eta_{\rm m}\rightarrow 1^{-}$,
which is in agreement with our expectations. The insert of Fig.~\ref{fig:var_1}(b)
illustrates the dependence of $\eta_{\rm m}(T_0)$ with the special cases
$T_0=T_{\rm c}(L)$ and $T_0\gg T_{\rm c}(L)$, which clearly
coincide with Eq.~(\ref{Eq:eta_m}).
Additional numerical study reveals the discrepancy for 
$T_0\ll T_{\rm c}(L)$, that is caused by a non-trivial
decrease of $\alpha$ by decreasing $T_0$ ($\alpha < 1$).
The $N$-dependence depicted in Fig.~\ref{fig:var_1}(b)
is constructed for an initial value $T_0=0.6$.
There is a certain $N$ value for which $\alpha=1$ 
and iterations cross from alternating to unidirectional
convergence. In other words, statistical errors
compensate systematic ones. In agreement with Ref.~\refcite{Munger91},
we have verified statistical errors of MC averages to be proportional
to~$1/\sqrt{N}$, which implies $A \sim 1/N$.
Actually, the fit provides $A \simeq 0.13/N$.

\begin{figure}[!ph]
\centerline{\epsfig{file=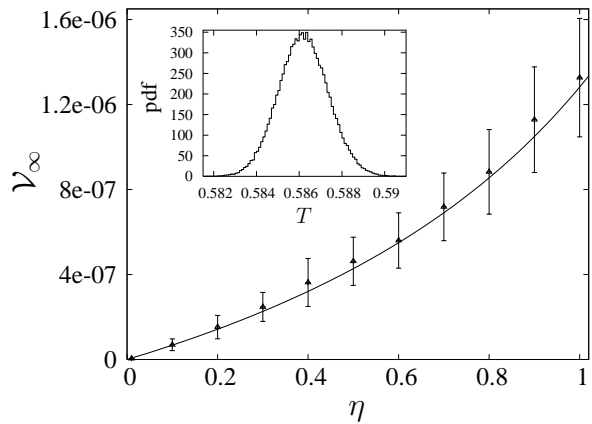,height=8cm,angle=-90.0}}
\caption{
         The $\eta$-dependence of
         ${\cal V}_{\infty}$
         compared with
         Eq.~(\ref{Eq:var_infty})
         for parameters $A$,
         $\alpha$ and $N=10^5$
         given in Table~\ref{tab:par}.
         The error bars
         are calculated using
         $5000$ independent runs.
         The inset depicts
         a stationary pdf of $T_t$
         constructed for
         $N_{\rm event}=10^5$
         and
         $\eta=1$.
\label{fig:var_infty}}
\end{figure}

If the number of iterations is much larger than $\tau_{\rm tr}$,
the initial conditions are forgotten and thus invariant
pdf's of temperature variable are formed~(see inset in
Fig.~\ref{fig:var_infty}). The calculation of
$T^{\ast}\equiv T_{\rm c}(L)$~($t\gg \tau_{\rm tr}\gg 1$)
is controlled by $\eta$. We have verified 
$\sqrt{{\cal V}_t}\propto \eta/\sqrt{N}$ for the {\em stationary regime}. 
The error increases monotonically with $\eta$ for a given $N$
as in Fig.~\ref{fig:var_infty}. We have compared
the simulation and analytical $\eta$-dependences
of ${\cal V}_{\infty}$ for $A$ and $\alpha$, given
in Table~\ref{tab:par}, and those determined by a fit of 
Eq.~(\ref{Eq:var_infty}) for $N=10^5$.
The comparison reveals a discrepancy of the $A$
parameter determination. The $A$ parameter, determined
by the fit of Eq.~(\ref{Eq:var_infty}) to numerical data,
increases its value about $6.2 \times 10^{-8}$.
A further extension of the model, e.g. taking into account
a weak colored noise, should eliminate this distinction.

According to~Eq.~(\ref{Tavel1}), the mean values of invariant 
pdf's~(see inset in Fig.~\ref{fig:var_infty})
were used to estimate $T_{\rm c}(10)=0.586141$.
The method is straightforward applied for $L=10,20,\ldots ,100$. 
Subsequently, the values $T_{\rm c}(L)$ 
are interpolated by the finite size formula
$T_{\rm c}(L)=T_{\rm c}+b/L$. This yields to estimate
the thermal coefficient $b=0.1889\pm 0.0033$
and the critical temperature as $T_{\rm c}=0.5673\pm 0.0001$.
The last estimation stays in a satisfactory agreement
with exact value $T_{\rm c}^{\rm ex}=
[2 \ln \left( 1 + \sqrt{2} \right)]^{-1}=0.56729$.
The problem of critical indices can be handled
effectively in the frame of a two-lattice
iterative algorithm, too~\cite{Horvath2004}.
In that case, the objective function given by Eq.~(\ref{Objec1})
has to be written in terms of the Binder
cumulant,
$\displaystyle
F:\, U_{T^{\rm his}_{t+1},T_t,N} = \min_{T\in\langle 0,\infty\rangle}
|U_{T,T_t,N}(L_2)-U_{T,T_t,N}(L_1)|$,
for two different system sizes.
According to the fact, that histogram reweighting allows
a superior determination of the derivation
$U_{T_{t+1},T_t,N}'(L) \equiv {\rm d} U_{T,T_t,N}(L)/{\rm d} T|_{T=T_{t+1}}$,
we considered, as an example, the exponent $\nu$ of the correlation length
\begin{equation}
\left(\frac{1}{\nu}\right)_t
\simeq
\frac
{\ln
\left[\displaystyle
      \frac{U_{T_{t+1},T_t,N}'(L_1)}{
            U_{T_{t+1},T_t,N}'(L_2)}
\right]}{\ln\left(L_1/L_2\right)}
\,.
\label{Eq:nu}
\end{equation}
Within the proposed algorithm, we generated a sequence
of estimates~(for each iteration step separately)
of the critical exponent for the lattice pair $2 L_1=L_2 = 20$.
Considering the peak of nonsymmetric pdf of $(1/\nu)_t$
we determined $1/\nu^{\rm ex}=1$ with a relative error of $1.35\%$.

\section{Conclusions}
An iterative process that is useful around the critical point and includes
Metropolis single spin-flip algorithm, single-histogram 
reweighting and linear filtering is studied numerically.
The properties of this iterative algorithm are studied for the 2d~Ising
ferromagnetic model on a square lattice.
The total errors in the estimate of pseudocritical temperature are obtained.
We demonstrated how the accuracy can be affected by~$\eta$.
However, one should be careful with an extremely small~$\eta$
to avoid an enormous increase of transient time.
Our analysis demonstrates that it is insufficient to use
several reweighting trials to estimate total errors.
The most interesting iterations are of this kind that
compensate the systematic errors introduced by the histogram 
reweighting approach.
Despite of the success of the linearized model, which allows
interpretation of numerical data, further improvement of 
the proposed methodology is still expected.
For instance, one should take into account non-linear
corrections in Eq.~(\ref{model_2}) and/or the colored 
noise~(that is, with autocorrelations of error terms) 
in~Eq.~(\ref{Eq:whitenoise}).
We believe that nontrivial dependences of $\alpha$ and
$\eta_{\rm m}$~(see e.g. inset of Fig.~\ref{fig:var_1}(b))
could be further explained considering microscopic
expression for the density of states~\cite{Beale96}.

\section*{Acknowledgments}
The authors would like to thank the Slovak Grant agency
VEGA~(grant No.~1/2009/05) and the agency APVT-51-052702 and
the internal grant of the Faculty of Sciences of \v{S}af\'arik
University VVGS 2003 for financial support. The authors express
their thanks to unknown referee for valuable corrections and 
suggestions.



\begin{thebibliography}{00}
\bibitem{Binder1988}
 K.~ Binder, D.W.~Heermann, {\it Monte Carlo Simulation in Statistical Physics}
 (Springer, Berlin, 1998).
\bibitem{Metropolis}
 N.~Metropolis, A.W.~Rosenbluth,
 M.N.~Rosenbluth, A.H.~Teller, {\it J. Chem. Phys.} {\bf 21}, 1087 (1953).
\bibitem{his_reweight}
 Z.W.~Salsburg, J.D.~Jacobson, W.~Fickett, W.W.~Wood, {\it J. Chem. Phys.} {\bf 30}, 65 (1959);\\
 A.M.~Ferrenberg, R.H.~Swendsen, {\it Phys. Rev. Lett.} {\bf 61}, 2635 (1988).
\bibitem{Ferrenberg63}
 A.M.~Ferrenberg, R.H.~Swendsen, {\it Phys. Rev. Lett.} {\bf 63}, 1195 (1989).
\bibitem{Ferrenberg95}
 A.M.~Ferrenberg, D.P.~Landau, R.H.~Swendsen, {\it Phys. Rev. E} {\bf 51}, 5092 (1995).
\bibitem{Ferrenberg91}
 A.M.~Ferrenberg, D.P.~Landau, {\it Phys. Rev. B} {\bf 44}, 5081 (1991).
\bibitem{Landau99}
 D.P.~Landau, {\it Journ. Magn. Magn. Mater.} {\bf 200}, 231 (1999).
\bibitem{Munger91}
 E.P.~M\"unger, M.A.~Novotny, {\it Phys. Rev. B} {\bf 43}, 5773 (1991).
\bibitem{Kadanoff}
 L.P.~Kadanoff, {\it Physics Today} (March 1991), p.~9.
\bibitem{Robbins51}
 H.~Robbins, S.~Munroe, {\it Ann. Math. Stat.} {\bf 22}, 400 (1951).
\bibitem{Ito94}
 N.~Ito, G.A.~Kohring, {\it Int. J. Mod. Phys. C} {\bf 5}, 1 (1994).
\bibitem{Plascak02}
 J.A.~Plascak, A.M.~Ferenberg, D.P.~Landau, {\it Phys. Rev. E} {\bf 65}, 066702 (2002).
\bibitem{Principe2000}
 J.C.~Principe, N.R.~Euliano, W.C.~Lefebvre, {\it Neural and adaptive systems:
 Fundamentals through simulations} (John Wiley \& Sons, Inc. 2000).
\bibitem{IntroductionBrockwell1996}
 P.J.~Brockwell, R.A.~Davis, {\it Introduction to Time Series and Forecasting}
 (Springer, New York, 1996).
\bibitem{Horvath2004}
 D.~Horv\'ath, M.~Gmitra, Z.~Kuscsik, {\it Czech. J. Phys.} {\bf 54}, 921 (2004);\\
 D.~Horv\'ath, M.~Gmitra, {\it Int. J. Mod. Phys. C} {\bf 15}, 1269 (2004).
\bibitem{Beale96}
 P.D.~Beale, {\it Phys. Rev. Lett.} {\bf 76}, 78 (1997).
\end{thebibliography}
\end{document}